\documentclass[seceq]{ptptex}

\usepackage{graphicx}
\usepackage{wrapft}

\title{
Accretion around supermassive black holes: 
the detection of the Balmer edge signature from quasars%
}

\author{
Makoto \textsc{Kishimoto}$^1$, 
Robert \textsc{Antonucci}$^2$, 
Catherine \textsc{Boisson}$^3$, 
Omer \textsc{Blaes}$^2$
}

\inst{
$^1$Institute for Astronomy, University of Edinburgh, 
Edinburgh EH9 3HJ, UK\\
$^2$Physics Department, University of California, Santa Barbara, CA
93106, USA\\
$^{3}$LUTH, FRE 2462 du CNRS, associ\'ee \`a l'Universit\'e Denis Diderot,
    Observatoire de Paris, Section de Meudon, F--92195 Meudon Cedex
}

\abst{

The ultraviolet/optical continuum of quasars is thought to be from an
accretion flow around a supermassive black hole, and it dominates the
radiative output of quasars. However, the nature of this emission, often
called the Big Blue Bump, has not been well understood.  Robust evidence
for its thermal nature would be a continuum opacity edge feature
intrinsic to this component, but this has not been clearly confirmed
despite the predictions by many models.  We are now developing and
exploiting a new method to detect the Balmer edge of hydrogen opacity.
The method overcomes for certain objects the major obstacle of the heavy
contamination from emissions outside the nucleus by taking the polarized
flux spectrum.  If our interpretation of the polarized flux is correct,
our data show that the Big Blue Bump emission has a Balmer edge in
absorption, indicating that the emission is indeed thermal, and the
emitter is optically thick.

}

\begin{document}

\maketitle

\section{introduction}

The ultraviolet/optical continuum radiation of quasars, which is often
called the Big Blue Bump, is thought to be from an accretion flow around
a supermassive black hole. The radiative output of quasars is dominated
by this component, and thus it is a crucial component to understand the
accretion flow.  This radiation is often assumed to be an optically
thick and thermal emission from an accretion disk.  However, continuum
edge features, which are key predictions of many models for the
atmospheric emission from accretion disks, have not been confirmed
satisfactorily.\cite{KB99,Bl01} \ Identifying these features would prove
the thermal nature of the emission in the first place, and thus it is of
fundamental importance.

There have been intensive and unsuccessful searches for broadened Lyman
edge discontinuities, starting with Ref.~\citen{An89}.  Later some slope
changes have been found around 1000\AA\ in some individual cases and
composite spectra\cite{Kr99,Zh97,Te02}.  This could be related to the
opacity change at the Lyman edge which is severely smeared by
relativistic effects.  However, in detail, even state-of-the-art
atmosphere models are not in a good agreement with the observed
feature.\cite{Bl01} \ In addition, a foreground absorption possibility
certainly complicates the interpretation of the spectral features in
this wavelength region.

The Balmer edge wavelengths can be more advantageous in these two
respects: (1) the spectral smearing would be less effective in the
context of accretion disk models, since this longer wavelength feature
originates from much farther out in the gravitational potential well;
(2) the Balmer edge wavelengths are less vulnerable to a foreground
absorption possibility since the Balmer features are not resonant ones
unlike the Lyman features. However, the emissions from outside the
nucleus, i.e. the broad-line region (BLR) and outer regions, heavily
contaminates this Balmer edge wavelength range. These are high-order
Balmer emission lines and Balmer continuum in emission as well as broad
FeII emission lines, collectively called the small blue bump.

There is a way to overcome this obstacle, however. Many normal quasars
show a small polarization ($P \sim 1$\%), which can be used at least in
some cases to eliminate any unwanted contamination from outside the
nucleus, as we describe below.

\section{Observations and results}

\begin{figure}[ht]

    \begin{minipage}{\textwidth}
       \includegraphics[width=6cm]{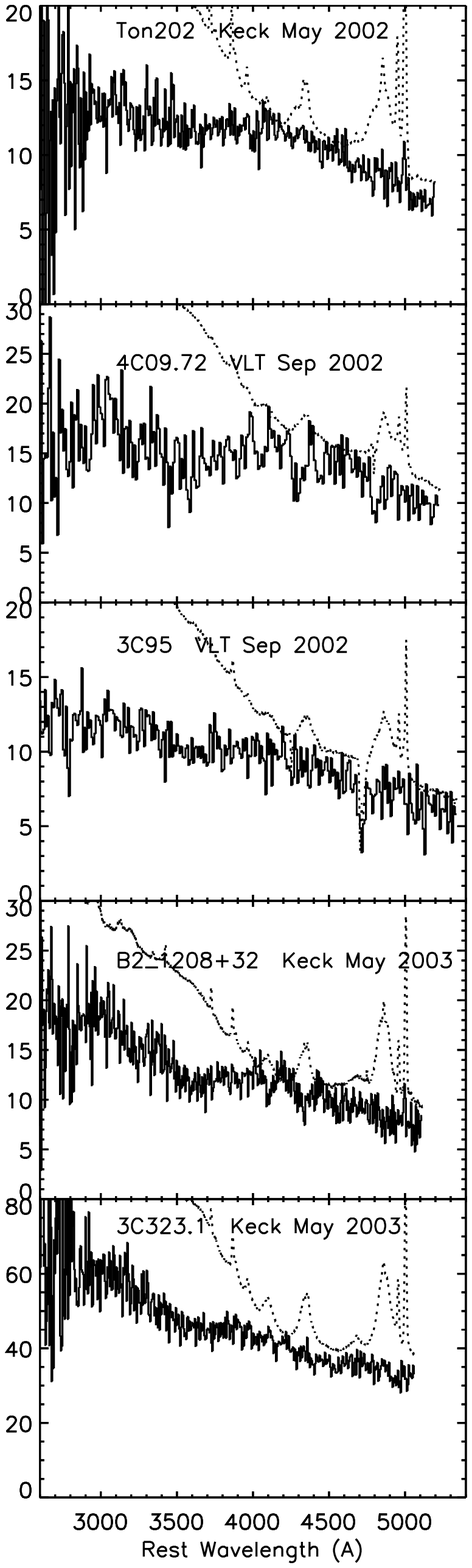}

       \caption{Our spectropolarimetric data for five quasars taken with
       the Keck and VLT.  The solid line represents the polarized flux,
       and the dotted line is the total flux scaled to roughly match at
       the red side. The left panel is in $F_{\lambda}$ and the right in
       $\nu F_{\nu}$ with both axes in log scale. The wavelengths are in
       the rest frame of each quasar. Note that the data for all the
       objects except Ton 202 have been corrected for the interstellar
       polarization in our Galaxy.\cite{KABB04} } \label{fig-5obj}
    \end{minipage}

    \vspace{-18.95cm} 
    \hspace{7cm} 
    \begin{minipage}{7cm}
       \includegraphics[width=6cm]{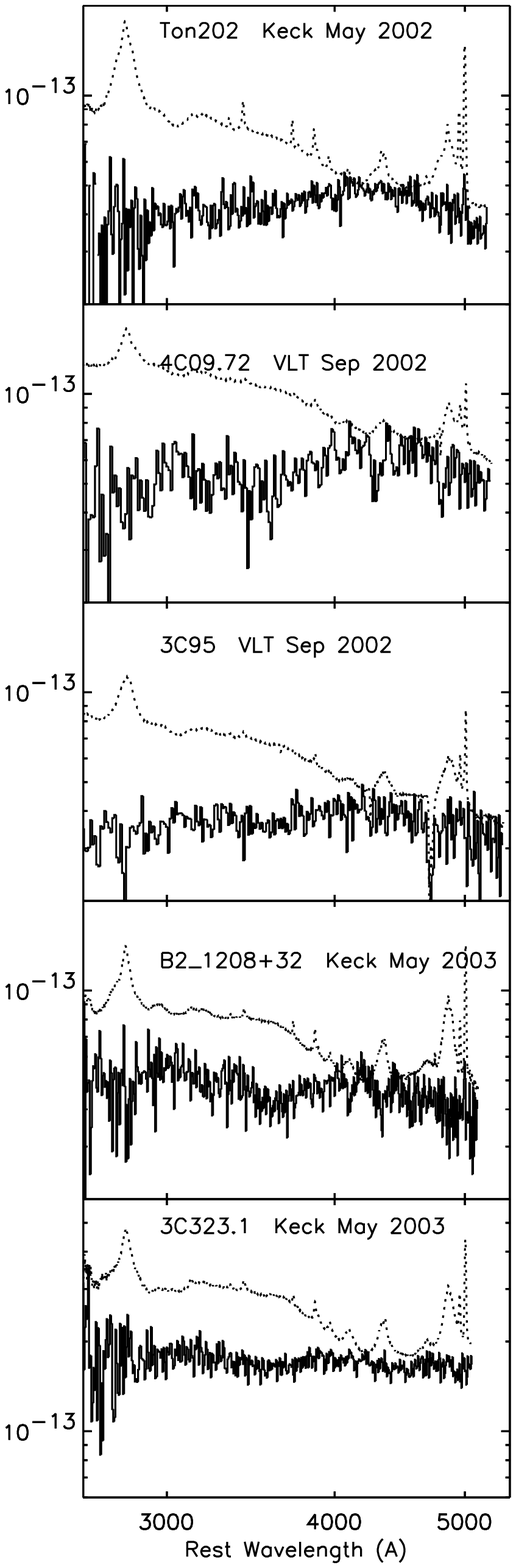}
    \end{minipage}

    \vspace{1.5cm}
\end{figure}

We have so far observed 16 quasars spectropolarimetrically with 8-10m
telescopes.  Firstly, two quasars were observed with the Keck telescope
in May 2002. The results have been published in Ref.~\citen{KAB03}
(Paper I). Then, 11 quasars were observed with the Very Large Telescope
(VLT) in September 2002 and three more new quasars were observed with
the Keck in May 2003 together with a re-observation of one of the
quasars observed in May 2002. The details of these VLT and second Keck
results will be published in Ref.~\citen{KABB04} (Paper II). Here, we
show the most favorable five objects for our Balmer edge investigation
in Fig.\ref{fig-5obj}, taken from Paper~II.  We note that the polarized
flux spectra of these five quasars were published in Ref.~\citen{SS00}
but with a lower S/N, so that the edge feature described below was not
clear.

The polarization spectra $P(\lambda)$ (not shown here) of these quasars
show a clear decrease at the broad emission line
wavelengths\cite{SS00,KAB03,KABB04}, so that essentially no emission
line shows up in the {\it polarized flux spectrum} $P \times
F_{\lambda}$ (note the distinction between $P$ and $P \times
F_{\lambda}$) to a high S/N as shown in Fig.\ref{fig-5obj}.  Just around
shortward of $\sim$4000\AA\ where the small blue bump emission starts in
the total flux spectrum, the polarization spectra again show a clear
decrease (not shown here), indicating that the small blue bump is also
unpolarized.\cite{SS00,KAB03,KABB04} \ Thus, the polarized flux is
essentially confined only to continuum, and the emission from the BLR
seems to be all unpolarized (Fig.\ref{fig-5obj}).

In all these five cases, this emission-line-free polarized flux seems to
show a Balmer edge feature in absorption.  The spectral slope of the
polarized flux at the wavelengths longward of $\sim$4000\AA\ is
essentially the same as that of the continuum in the total flux, but at
$\sim$4000\AA, there is a slope down-turn, which is considered to be the
start of the Balmer edge feature.  Then, there also seems to be a slope
up-turn at $\sim$3600\AA.  The edge feature seems to be broadened
possibly due to a rather high velocity dispersion.

\section{Discussion and conclusion}

The continuum-confined polarization seen in these objects is quite in
contrast to Seyfert 2 galaxies where the continuum and the broad
emission lines are {\it both} polarized.  Based on this, the
polarization is thought to occur due to scattering {\it outside} the
BLR.\cite{AM85} \ However, in our cases, the emission from the BLR is
not polarized. Therefore, the polarization is considered to occur {\it
interior} to the BLR. Thus, the polarized flux is likely to show the
intrinsic spectral shape of the emission from interior to the BLR,
scraping off all the other contaminating emission from outside.

The cause of the polarization is not still clear, but probably it
involves some form of electron scattering. Dust scattering is not
likely, since the region which is causing the polarization is thought to
be interior to the BLR and thus within the dust sublimation radius.  The
scattering region could be the atmosphere of the putative accretion disk
around the central black hole (though see more discussions in
Paper~I). Alternatively, it could be a diffuse optically-thin region
surrounding the emitter of the Big Blue Bump.  In either of these
simplest cases, the Balmer edge feature seen in absorption in the
polarized flux is intrinsic to the Big Blue Bump emission. In this case,
the feature indicates that the emission is indeed thermal, and the
emitter is optically thick.

However, some alternative explanations for the observed polarized flux
are not ruled out, as discussed in Paper I. Briefly, the feature might
have been imprinted in some foreground region or in the
scattering/polarizing region itself. However, a simple foreground
absorption is unlikely due to the general lack of a huge, corresponding
Lyman edge absorption in the total flux.  If the Balmer edge feature
were to be imprinted in-situ when the polarized flux is formed in the
scattering region outside the Big Blue Bump emitter, then the
distinction from the case of the atmospheric feature might be just
semantic, since the scattering region should be optically thick, and
should have a large velocity dispersion to produce a large broadening in
the edge feature.

In the re-observation of Ton 202 in May 2003, which is one year apart
from the observation shown in the top panels of Fig.\ref{fig-5obj}, we
did not detect such a dramatic feature. The details are presented in
Paper II, but this is apparently due to polarization variability.  This
is not unexpected, since, as we discussed above, the polarization is
thought to originate from a quite compact region, at least comparable to
or smaller than the size scale of the BLR.

In conclusion , the edge feature is most simply interpreted as an
intrinsic feature of the Big Blue Bump emission, showing its thermal and
optically thick nature. However, we cannot rule out some alternative
interpretations, and we need more data and modeling to confirm that 
our interpretation is correct.

\section*{Acknowledgements}
MK appreciates the generous financial support from the conference
organizers. The work by RA was supported in part by NSF grant
AST-0098719.

%

\end{document}